\documentclass[aps,pra,twocolumn]{revtex4}

\usepackage{amsmath}
\usepackage{amssymb}
\usepackage{mathrsfs}
\usepackage{epsfig}
\usepackage{color}
\usepackage[bookmarks]{hyperref}
\usepackage{filecontents}
\usepackage{graphicx}
\usepackage{mathrsfs}
\usepackage{dsfont}
\graphicspath{{.}{./}}
\usepackage{enumerate}
\usepackage{mathtools}
\usepackage{bm}

\usepackage{soul}

\newcommand{\eref}[1]{Eq.~(\ref{eq:#1})}

\newcommand* {\bra}[1]{\ensuremath{\langle {#1} |}}
\newcommand* {\ket}[1]{\ensuremath{| {#1} \rangle}}

\begin{document}
\title{Fisher-information-based estimation of optomechanical coupling strengths}

\author{Claudio Sanavio}
\email{claudio.sanavio@um.edu.mt}
\affiliation{Department of Physics, University of Malta, Msida MSD 2080, Malta}
\author{J\'ozsef Zsolt Bern\'ad}
\affiliation{Department of Physics, University of Malta, Msida MSD 2080, Malta}
\author{Andr\'e Xuereb}
\affiliation{Department of Physics, University of Malta, Msida MSD 2080, Malta}

\date{\today}

\begin{abstract}
The formalism of quantum estimation theory, focusing on the quantum and classical Fisher information, is applied to the estimation of the coupling strength in an optomechanical system. In order to estimate the optomechanical coupling, we 
have considered a cavity optomechanical model with non-Markovian Brownian motion of the mirror and employed input--output formalism to obtain the cavity output field. Our estimation scenario is based 
on balanced homodyne photodetection of the cavity output field. We have explored the difference between the associated measurement-dependent classical Fisher information and the quantum 
Fisher information, thus addressing the question of whether it is possible to reach the lower bound of the mean squared error of an unbiased estimator by means of balanced homodyne detection. We have found that the phase of the local oscillator in the homodyne
detection is crucial; certain quadrature measurements allow very accurate estimation.
\end{abstract}

\maketitle

\section{Introduction}

Inverse problems play an important role in science because they are able to inform us about relevant parameter values of a dynamical system that we cannot directly observe \cite{Kaipio}. The objective of an 
inverse problem 
is to estimate these unknown parameters by extracting information from measurement data and assessing the uncertainty in these data, making use of all information known prior to the measurement process and a 
mathematical model of the dynamical system. 
In this approach, the parameters to be estimated are treated as random variables and they must be assigned a joint prior probability distribution function; this is the Bayesian formulation of the 
estimation problem. The qualities of estimators acting on the space of measurement data are evaluated through cost functions or, conversely, by maximizing or minimizing a cost function over the set of all possible 
estimators that leads to an optimal estimator. In this case, calculus of variations is applied, which is not always an easy mathematical task, especially when the estimation problem is 
formulated in quantum mechanics \cite{Personick, Holevo, Helstrom}. Applications to quantum mechanical systems do not always result in an experimentally implementable optimal estimator 
\cite{Macieszczak, Rzadkowski, Bernad1, Rubio, Bernad2}. 

Optimal estimators, in general, are likely to be complicated, as is observed in our previous investigation on the estimation of the nonlinear optomechanical coupling strength \cite{Bernad1}. Furthermore, solving the 
variational problem for the average cost function imposes limits on the use of models with many types of decoherence sources. In order to work with more effective models of cavity optomechanical systems \
\cite{Aspelmeyer2014} and to consider experimentally relevant estimation strategies \cite{Brawley2016}, one has to turn to the investigation of the lower bounds of some convenient measure of the estimation accuracy. 
The mean-squared error- the average squared difference between estimated values and true values of the unknown parameters- is usually employed as a measure of accuracy. In the case of classical systems, 
there are some complicated lower bounds of the mean-squared error \cite{Bhattacharyya, Barankin}; however, the Cram\'{e}r--Rao inequality \cite{Rao, Cramer}, which defines an inferior but a simpler lower bound, 
can be extended to quantum systems \cite{Helstrom2}. Here, the lower bound is inversely proportional to the quantum Fisher information (QFI), irrespective of whether the estimator is biased or unbiased; see Ref. 
\cite{Bernad2}. The chosen estimation strategy, expressed as a positive-operator valued measure (POVM), provides probability distributions of the parameter to be estimated,
conditioned on the true value of this parameter. These probabilities determine the classical Fisher information (CFI), which is inversely proportional to the lower bound of the mean-squared error in
the classical post-processing of measurement data. As the CFI is always smaller than or equal to the QFI, which defines the smallest value of the lower bound, 
it is worthwhile to investigate the circumstances where the CFI is as close as possible to the QFI \cite{Matteo}. If the QFI is saturated by some POVM and the probability distributions 
belong to a one-parameter exponential family, which is a requirement that our theoretical approach fulfills, then there exists a suitable classical unbiased estimator on the measurement data \cite{Casella}, 
which saturates the CFI and thus yields the most precise measurement.

In this paper, we follow the above-described methodology, which allows us to consider a detailed model of a cavity optomechanical system. We consider a single mode of the radiation field inside a cavity and 
also a single vibrational mode of the mechanical resonator. The two modes interact via a radiation-pressure interaction Hamiltonian \cite{Law}. 
The single-mode field assumption is justified when the cavity is driven by an external laser with a bandwidth significantly narrower than the separation among 
the different electromagnetic field modes. The laser populates only one mode, allowing us to neglect the others. There are many mechanical modes, but describing only one of them has proven to 
be a valid approximation in experiment \cite{Aspelmeyer2014}.
The mechanical oscillator is subject to quantum Brownian motion \cite{CL, Breuer} and the single-mode cavity field is coupled to the electromagnetic field outside of the cavity \cite{Walls}. 
Finally, balanced homodyne photodetection with non-ideal detectors \cite{Raymer} is carried out on the output field; this automatically defines the set of POVMs, i.e., the estimation strategy, that we explore. 
We investigate the QFI of the output field state depending on the unknown value of the nonlinear optomechanical coupling and 
compare with the CFI obtained from the data provided by the balanced homodyne photodetection. We identify those cases where CFI is as large as possible, where the lower bound of the estimation 
accuracy is therefore smallest.  

This paper is organized as follows. In Sec. \ref{I}, we discuss the cavity optomechanical model and determine the stationary state of the output field.  
In Sec. \ref{II}, the QFI of the output field state is determined. A brief overview of balanced homodyne photodetection and the related POVM is presented and probabilities 
of these measurements are given, which allows us to calculate the CFI. A numerical investigation and the maximization of CFI with respect to QFI are addressed in Sec. \ref{III}. 
Finally, in Sec. \ref{IV}, we draw conclusions and make some remarks on our work. Detailed derivations supporting the main text are collected in the two appendices.

\section{Model}
\label{I}

The optomechanical system we have in mind is formed by a Fabry--P\'{e}rot cavity with a moving-end mirror and we focus on a case where 
only a single mode of the radiation field and a vibrational mode of the mechanical oscillator, i.e., moving mirror, are considered. The model can be used to describe several alternative systems \cite{Aspelmeyer2014}.
The free Hamiltonian of the system reads
\begin{equation}\label{freeHamiltonian}
\hat{H}_0=\hbar\omega_{\text{c}}\hat{a}^\dagger \hat{a}+\frac{\hat{p}^2}{2m}+\frac{1}{2}m\omega_{\text{m}}^2\hat{q}^2,
\end{equation} 
where $\hat{q}$ and $\hat{p}$ are position and momentum operators for the mechanical oscillator of effective mass $m$ and which oscillates at frequency $\omega_{\text{m}}$. The annihilation and 
creation operators of the single-mode radiation field with frequency $\omega_{\text{c}}$ are denoted by $\hat{a}$ and $\hat{a}^\dagger$.
The two subsystems are coupled by the optomechanical interaction \cite{Law}, which is the radiation pressure on the oscillating mirror, which is well described by a non--linear Hamiltonian term \cite{Aspelmeyer2014},
\begin{equation}\label{OptomechanicalInteraction}
\hat{H}_I=-\hbar g\hat{a}^\dagger \hat{a}\hat{q},
\end{equation}
with coupling strength $g$. The use of the term ``non--linear'' here refers to the equation of motion of the system operators, at least one of which is non--linear, and is related to the Hamiltonian being of third order in these operators. 

In order to describe this optomechanical system effectively, one has to consider decoherence and excitation losses, i.e., the concept of open quantum systems has to be applied.
The single-mode field is affected by a decay with rate $\kappa=\kappa_{\text{in}}+\kappa_{\text{loss}}$, where $\kappa_{\text{in}}$ is the loss rate associated with the input--output fields and $\kappa_{\text{loss}}$ is related to what are commonly called internal 
losses \cite{Tagantsev}. The latter quantity could, for example, originate from the fact that the cavity mirrors act to scatter photons from the cavity mode of interest to other modes or to the outside environment. The mechanical oscillator is in 
contact with a phonon bath at temperature $T$ and experiences a friction or decay rate $\gamma$. The dynamics is given in the Heisenberg picture with the use of the quantum Langevin equations,
\begin{eqnarray}\label{dadt}
\dot{\hat{a}}&=&-i\omega_{\text{c}} \hat{a}+ig\hat{a}\hat{q}-\frac{\kappa}{2}\hat{a}+\sqrt{\kappa_{\text{in}}}\hat{a}_{\text{in}}+\sqrt{\kappa_{\text{loss}}}\hat{a}_{\text{loss}},\\\label{dadaggerdt}
\dot{\hat{a}}^\dagger&=&i\omega_{\text{c}}\hat{a}^\dagger-ig\hat{a}^\dagger\hat{q}-\frac{\kappa}{2}\hat{a}^\dagger+\sqrt{\kappa_{\text{in}}}\hat{a}^\dagger_{\text{in}}+\sqrt{\kappa_{\text{loss}}}\hat{a}^\dagger_{\text{loss}},\\\label{dqdt}
\dot{\hat{q}}&=&\frac{\hat{p}}{m},\\\label{dpdt}
\dot{\hat{p}}&=&-m\omega_{\text{m}}^2\hat{q}-\gamma\hat{p}+\hbar g\hat{a}^\dagger\hat{a}+\hat{\xi}.
\end{eqnarray}
where $\hat{a}_{\text{in}}$ is the input noise operator associated with the modes of the radiation field outside the
cavity. $\hat{a}_{\text{loss}}$ is the operator describing the internal losses and $\hat{\xi}$ represents the quantum Brownian noise operator. 
Making use of the spectral density $J(\omega)$ of the phonon modes in the bath and the weak coupling of the mechanical oscillator to the bath \cite{Breuer-Kappler},
one can define the following functions \cite{GiovVit}:
\begin{eqnarray}\label{Dreal}
\mathfrak{D}_R(\tau)&=&\int^\infty_0 d\omega J(\omega)\cos(\omega\tau)\coth\bigg{(}\frac{\hbar \omega}{2 k_BT}\bigg{)}\\\label{Dimaginary}
\mathfrak{D}_I(\tau)&=&\int^\infty_0 d\omega J(\omega)\sin(\omega\tau).
\end{eqnarray}
Now, we are able to calculate the two-time correlation function of $\hat{\xi}(t)$,
\begin{eqnarray}
\langle\hat{\xi}(t)\hat{\xi}(t')\rangle=\hbar[\mathfrak{D}_R(t-t')-i\mathfrak{D}_I(t-t')].
\end{eqnarray}
The mean of $\hat{\xi}(t)$ is zero and its non-Markovian nature allows us to preserve the correct
commutation relations between $\hat{p}$ and $\hat{q}$ during the time evolution \cite{GiovVit}. An extensively studied case is the ohmic spectral density with a Lorentz--Drude cutoff function,
\begin{equation}
 J(\omega)=\frac{2 m \gamma}{\pi} \omega \frac{\Omega^2}{\omega^2 + \Omega^2}, \nonumber
\end{equation}
where $\Omega$ is the high-frequency cutoff. An ohmic spectral density with exponential cutoff \cite{Garg}, 
\begin{equation}\label{Ohmicspectraldensity}
J(\omega)=\frac{2 m \gamma}{\pi} \omega \exp\bigg{(}{-\frac{\omega}{\Omega}}\bigg{)}
\end{equation}
leads to very similar behavior to one with a Lorentz--Drude cutoff function, albeit with the advantage that the integrations
in Eqs. \eqref{Dreal} and \eqref{Dimaginary} have analytical solutions in closed form,  
\begin{eqnarray}
\mathfrak{D}_R(\tau)&=&\frac{2 m \gamma}{\pi} \frac{\Omega^2 \left(\Omega^2 \tau^2-1 \right)}{\left(\Omega^2 \tau^2+1 \right)^2 } + \nonumber \\
&& \frac{2 m \gamma}{\pi \hbar^2} \left(k_B T\right)^2 \left[\Psi^{(1)}(z)+\Psi^{(1)}(z^*) \right], \nonumber \\
z&=&\frac{1-i\Omega \tau}{\hbar \Omega} k_B T, \nonumber \\
\mathfrak{D}_I(\tau)&=& \frac{2 m \gamma}{\pi} \frac{2 \Omega^3 \tau}{\left(\Omega^2 \tau^2+1\right)^2} \nonumber
\end{eqnarray}
with $\Psi^{(1)}(z)$ being the polygamma function \cite{Stegun}. The cavity operates at optical frequencies, i.e., $\hbar \omega _c /k_B T \gg 1 $ holds to a very good approximation at reasonable temperatures, 
and therefore the operators 
$\hat{a}_{\text{in}}(t)$ and $\hat{a}^\dagger_{\text{in}}(t')$ commute for $t \neq t'$. Their correlation functions in the vacuum state $\ket{0}$ read 
\begin{eqnarray}
 \bra{0} \hat{a}_{\text{in}}(t) \hat{a}^\dagger_{\text{in}}(t') \ket{0}=\delta(t-t'), \nonumber \\
 \bra{0} \hat{a}^\dagger_{\text{in}}(t') \hat{a}_{\text{in}}(t) \ket{0}=0. \nonumber
\end{eqnarray}
The operators $\hat{a}_{\text{loss}}(t)$ and $\hat{a}^\dagger_{\text{loss}}$ have similar commutation relations and, furthermore, they commute at all times with  
$\hat{a}_{\text{in}}(t)$ and $\hat{a}^\dagger_{\text{in}}(t')$.

Usually the single mode of the cavity is driven by a laser with frequency $\omega_{\text{L}}$ and intensity $\epsilon$. This process can be modified through the addition of the following term to the Hamiltonian:
\begin{equation}\label{laserHamiltonian}
H_{\text{L}}=i\hbar\epsilon\left(\hat{a}^\dagger e^{-i\omega_{\text{L}}t}-\hat{a}e^{i\omega_{\text{L}}t}\right),
\end{equation}
whose phases $\pm \omega_{\text{L}} t$ can be easily absorbed after going into a rotating frame, with a resulting detuning for the optical frequency $\Delta_0=\omega_{\text{c}}-\omega_{\text{L}}$. 
In terms of the power $P$ of the laser, the driving intensity is $\epsilon=\sqrt{2\kappa_{\text{in}}P/\hbar \omega_{\text{L}}}$.

It is worth noting that the driving of the field is a necessary condition to obtain an effectively linear optomechanical interaction \cite{Aspelmeyer2014}. 
The application of a high-intensity laser field causes the single-mode field to reach a steady state with finite amplitude $\alpha$ ($|\alpha| \gg 1$) 
and allows us to consider only the quantum fluctuations around this stationary state. This also affects the mechanical oscillator by shifting the minimum of the harmonic potential. 
The dynamics of the fluctuations around the steady state is well described by linearizing the quantum Langevin equations \eqref{dadt}--\eqref{dpdt}. 
This can mathematically  be described by the application of two displacement operators ,
\begin{equation}
 \hat{D}_1 (\alpha)=e^{\alpha \hat{a}^\dagger-\alpha^* \hat{a}} \quad \text{and} \quad \hat{D}_2 (\beta_0)=e^{\beta_0 \hat{b}^\dagger-\beta^*_0 \hat{b}} \nonumber 
\end{equation}
with
\begin{eqnarray}
\hat{b} & = & \sqrt{\frac{m\omega_{\text{m}}}{2\hbar}}\bigg{(}\hat{q}+\frac{i}{m\omega_{\text{m}}}\hat{p}\bigg{)}\\
\hat{b}^{\dagger} & = & \sqrt{\frac{m\omega_{\text{m}}}{2\hbar}}\bigg{(}\hat{q}-\frac{i}{m\omega_{\text{m}}}\hat{p}\bigg{)}
\end{eqnarray}
on the quantum Langevin equations \eqref{dadt}--\eqref{dpdt}. The above equations also define relations between $(\beta_0, \beta^*_0)$ and $(q_0, p_0)$.
A transformation back to the operators $\hat{p}$ and $\hat{q}$, with the quantities introduced above satisfying the non--linear equation
\begin{equation}
i\epsilon=-\Delta_0\alpha+ g\alpha q_0-i\frac{\kappa}{2}\alpha \label{conditionLaser}
\end{equation}
yields a driving free Hamiltonian,
\begin{eqnarray}\label{fullHamiltonian}
\hbar\Delta\hat{a}^\dagger \hat{a}+\frac{\hat{p}^2}{2m}+\frac{1}{2}m\omega_{\text{m}}^2\hat{q}^2-\hbar g\alpha(\hat{a}+\hat{a}^\dagger)\hat{q},
\end{eqnarray}
where
\begin{eqnarray}\label{conditionDelta}
\Delta&=&\Delta_0-\hbar\frac{g^2|\alpha|^2}{m \omega^2_m},\\\label{conditionshift}
q_0&=&\hbar\frac{g|\alpha|^2}{m \omega^2_m}.
\end{eqnarray}

Equation \eqref{conditionLaser}, together with the above equations, yields a third degree equation for $|\alpha|$. Depending on the value of the power $P$, we may encounter a bistability of the system that will give two 
different solutions for the shift in the rest position of the mirror \eqref{conditionshift}. One should note that the steady-state amplitude depends on the value of $g$, making $\alpha$ a function of $g$, i.e. $\alpha=\alpha(g)$. 
The same is valid for the detuning $\Delta=\Delta(g)$. Therefore, for a fixed value of $P$, the bistability also depends explicitly on $g$, which is as yet unknown. A good strategy here is to define an interval, 
depending on our prior knowledge, for the possible values of $g$ and adjust the power of laser $P$ such that the bistability is completely avoided. For detailed calculations and regions of 
stability and bistability, see Appendix \ref{sec:laser}.

The shifted operators (also denoted fluctuation operators) $\delta \hat{a}=\hat{a}-\alpha $ and $\delta \hat{q}=\hat{q}-q_0$ are subject to the same loss process as the original ones. Note that the momentum operator $\delta\hat{p}=\hat{p}$ is 
not changed because $\beta_0$ is real, which implies that $p_0=0$. It is more convenient to define the two quadratures of the single-mode field $\delta \hat{X}=(\delta \hat{a}^\dagger+\delta \hat{a})/\sqrt{2}$ and 
$\delta \hat{Y}=i(\delta \hat{a}^\dagger-\delta \hat{a})/\sqrt{2}$. We analogously define the quadratures $\hat{X}_{\text{in}}$, $\hat{Y}_{\text{in}}$, $\hat{X}_{\text{loss}}$, and $\hat{Y}_{\text{loss}}$. 
Under the assumption that $\lvert\alpha\rvert$ is large, we can truncate the equations of motion to first order in the fluctuation operators. Finally, the differential equations of the shifted operators can be written in the concise form
\begin{equation}\label{uevolution}
\dot{u}(t)=Au(t)+\eta(t),
\end{equation}
where we have defined the vector of operators $u(t)=\left(\delta \hat{q}(t),\delta \hat{p}(t),\delta \hat{X}(t),\delta \hat{Y}(t)\right)^T$ and
\begin{eqnarray}
 \eta(t)=&&\left(0,\hat{\xi}(t),\sqrt{\kappa_{\text{in}}}\hat{X}_{\text{in}}(t)+\sqrt{\kappa_{\text{loss}}}\hat{X}_{\text{loss}}(t), \right. \nonumber \\
 &&\left. \sqrt{\kappa_{\text{in}}} \hat{Y}_{\text{in}}(t)+\sqrt{\kappa_{\text{loss}}} \hat{Y}_{\text{loss}}(t) \right)^T \nonumber;
\end{eqnarray}
and the superscript $T$ denotes the transposition. Furthermore, we have
\begin{equation}\label{eq:Amatrix}
A=\begin{pmatrix}
 0 & \frac{1}{m} & 0 & 0 \\
 -m \omega _{\text{m}}^2 & -\gamma  & \sqrt{2}  \hbar g \alpha(g)  & 0 \\
 0 & 0 & -\frac{\kappa }{2} & \Delta(g)  \\
 \sqrt{2} g \alpha(g)  & 0 & -\Delta(g)  & -\frac{\kappa }{2}, 
 \end{pmatrix}
\end{equation}
where we have introduced the explicit dependence of $\alpha(g)$ and $\Delta(g)$ on $g$.
The solution to \eqref{uevolution} reads
\begin{equation}\label{usolution}
u(t)=\exp(A t)u(0)+\int_0^t dt'\,\exp[A (t-t')]\eta(t').
\end{equation}
The autocorrelation matrix is given by
\begin{equation}
 R(t,s)=\left \langle u(t) u(s)^T \right \rangle. \nonumber
\end{equation}
Making use of the relation
\begin{equation}
\left \langle u(0) \eta(t)^T \right \rangle=\left \langle \eta(t) u(0)^T \right \rangle=0, \quad t\geqslant 0, \nonumber 
\end{equation}
one finds
\begin{eqnarray}
 &&R(t,s)=\exp(A t) \left \langle u(0) u(0)^T \right \rangle\exp(A^T s) \nonumber \\
 &&+ \int^t_0 \int^s_0 dt' dt'' \exp[A (t-t')] M(t'-t'') \exp[A^T (s-t'')] \nonumber
\end{eqnarray}
where
\begin{eqnarray}
 &&M(t'-t'')=\left \langle \eta(t') \eta(t'')^T \right \rangle \nonumber \\ 
 &&=\begin{pmatrix}
 0 & 0 & 0 & 0 \\
 0 & \langle\hat{\xi}(t')\hat{\xi}(t'')\rangle  & 0  & 0 \\
 0 & 0 & \frac{\kappa}{2} \delta(t'-t'') & i \frac{\kappa}{2} \delta(t'-t'')  \\
 0 & 0 & -i \frac{\kappa}{2} \delta(t'-t'')   & \frac{\kappa}{2} \delta(t'-t'')
 \end{pmatrix}.\nonumber
\end{eqnarray}
Let us consider the symmetric autocorrelation matrix
\begin{equation}
 \sigma(t,s)=\frac{R(t,s)+R^T(t,s)}{2}. \nonumber
\end{equation}
Taking $t=s$ we obtain
\begin{equation}
 \frac{d}{dt}\sigma(t)=A \sigma(t)+\sigma(t) A^T+D(t), \nonumber
\end{equation}
where $\sigma(t)\equiv\sigma(t,t)$, and with
\begin{eqnarray}
 D(t)&=&\int^t_0  dt' \frac{M(t-t')+M^T(t-t')}{2} \exp[A^T (t-t')] \nonumber \\
 &+&\int^t_0 dt' \exp[A (t-t')] \frac{M(t'-t)+M^T(t'-t)}{2}. \nonumber
\end{eqnarray}
This can be further simplified via
\begin{eqnarray}
 M_1(t-t')&=&\frac{M(t-t')+M^T(t-t')}{2} \nonumber \\
 &=&\begin{pmatrix}
 0 & 0 & 0 & 0 \\
 0 & \hbar \mathfrak{D}_R(t-t')  & 0  & 0 \\
 0 & 0 & \frac{\kappa}{2} \delta(t-t') & 0 \\
 0 & 0 & 0  & \frac{\kappa}{2} \delta(t-t')
 \end{pmatrix}, \nonumber
\end{eqnarray}
because $\mathfrak{D}_R(-t)=\mathfrak{D}_R(t)$ and $\mathfrak{D}_I(-t)=-\mathfrak{D}_I(t)$, which also implies
\begin{equation}
 M_1(t-t')=M_1(t'-t). \nonumber
\end{equation}
Finally, we can write
\begin{equation}
 D(t)=\int^t_0 d\tau \Big[ M_1(\tau) \exp(A^T \tau)+ \exp(A \tau)  M_1(\tau) \Big]. \nonumber
\end{equation}

The stability of the system, $\lim _{t \to \infty} \exp(A t)=0$, can be derived by applying the Routh--Hurwitz
criterion \cite{Gant}. This has been thoroughly investigated in the last decade and the two nontrivial conditions on the parameters of $A$ show that if the system is stable, then the bistability 
of the dynamics is avoided \cite{Genes2008}. From now on, we consider these conditions to be satisfied.
Therefore, $u(t)$ for $t \to \infty$ approaches zero, which implies that the autocorrelation matrix $\sigma(t)$ coincides with the matrix in the stationary solution. 
The stationary correlation matrix is defined as $\sigma=\lim_{t \to \infty} \sigma(t,t)$ and is the solution to the 
following Lyapunov equation:
\begin{equation}\label{Lyapunov}
A\sigma+\sigma A^T=-D,
\end{equation}
where
\begin{equation}
D =\int_0^{\infty} d\tau \Big[ M_1(\tau) \exp(A^T \tau)+ \exp(A \tau)  M_1(\tau) \Big]. \nonumber
\end{equation}

We need to keep in mind that any experimental apparatus does not have direct access to the cavity field, but only to the output field, which 
escapes the cavity. We can calculate the fluctuations of this field around its stationary state with the use of the input--output relations,
\begin{equation}\label{inputoutput}
\hat{a}_{\text{out}}=\sqrt{\kappa_{\text{in}}}\delta \hat{a}-\hat{a}_{\text{\text{in}}}.
\end{equation}
In practice, one selects different modes by opening a filter in a certain interval of time or in different
frequency intervals. Hence, we can define $N$-independent output modes following the approach of Ref.~\cite{Genes2008},
\begin{equation}\label{outputmodes}
\hat{a}_{k,{\text{out}}}(t)=\int_{-\infty}^t ds g_k(t-s)\hat{a}_{\text{out}}(s),\quad k=1,\dots,N,
\end{equation}
where $g_k(s)$ is the filter function defining the $k\textsuperscript{th}$ mode. Here we will make use of the filter function,
\begin{eqnarray}\label{eq:filter}
g_k(t)=\frac{\theta(t)-\theta(t-\tau)}{\sqrt{\tau}}e^{-i\Omega_kt},
\end{eqnarray}
with $\Omega_j-\Omega_l=\frac{2\pi}{\tau}n,$ $n\in\mathbb{N}$. The $k\textsuperscript{th}$ mode is centered at the frequency $\Omega_k$ and has a bandwidth $1/\tau$.
Making use of the input--output relations \eqref{inputoutput}, we obtain the correlation matrix $\sigma_{k,{\text{out}}}$ 
of the output field quadratures $\hat{X}_{k,\text{out}}$ and $\hat{Y}_{k,\text{out}}$ related to a filter centered at frequency $\Omega_k$ as (see Appendix \ref{sec:sigmakout})
\begin{widetext}
\begin{eqnarray}\label{correlationmatrixoutput1}
\langle\hat{X}_{k,\text{out}} \hat{X}_{k,\text{out}}\rangle(\tau)&=&\frac{1}{2} \kappa  \tau  \text{sinc}^2\left(\frac{\Omega_k \tau}{2}\right) \left[\left(\sigma_{xx}-\sigma_{yy}\right) 
\cos (\Omega_k \tau)+\sigma_{xx}+2 \sigma_{xy} \sin (\Omega_k \tau)+\sigma_{yy}\right]+\text{sinc}\left(2 \Omega_k \tau \right)  \\ \label{correlationmatrixoutput2}
\langle\hat{X}_{k,\text{out}}\hat{Y}_{k,\text{out}}\rangle(\tau)&=&\frac{1}{2} \kappa  \tau  \text{sinc}\left(\frac{\Omega_k \tau}{2}\right)^2 \left[ \left(\sigma_{yy}-\sigma_{xx}\right) 
\sin ( \Omega_k \tau )+2
  \sigma_{xy} \cos (\Omega_k \tau) \right]\\\label{correlationmatrixoutput3}
\langle\hat{Y}_{k,\text{out}}\hat{Y}_{k,\text{out}}\rangle(\tau)&=&\frac{1}{2} \kappa  \tau  \text{sinc}^2\left(\frac{\Omega_k \tau}{2}\right) \left[ \left( \sigma_{yy}- \sigma_{xx}\right) 
\cos (\Omega_k \tau)+ \sigma_{xx}-2  \sigma_{xy} \sin (\Omega_k \tau)+ \sigma_{yy}\right]+\text{sinc}\left(2 \Omega_k \tau \right)
\end{eqnarray}
\end{widetext}
where $\sigma_{AB}=\left \langle \delta \hat{A} \delta \hat{B}\right \rangle $ ($A,B \in \{X,Y\}$) are the entries of matrix $\sigma$, which are obtained in Eq. \eqref{Lyapunov}, and 
$\text{sinc}(x)$ is the unnormalized sinc function $\text{sinc}(x)=\sin(x)/x$.
The shifted operators are fully characterized in the stationary state by the correlation matrix, since all noises involved obey this property and the equations of motion are linear. One can thus deduce that
their properties can also be described by a zero-mean  Gaussian state. Similarly, the output field fluctuations are given by the Gaussian Wigner function
\begin{equation}\label{Wigner}
W(\xi)=\frac{1}{\sqrt{2\pi\det(\sigma_{k, \text{out}})}}e^{-\frac{1}{2}\xi^T \sigma^{-1}_{k,\text{out}} \xi},
\end{equation}
where $\xi=(X_{k,\text{out}},Y_{k, \text{out}})^T$ and $\sigma_{k, \text{out}}$ is the correlation matrix. The above Wigner function depends on the optomechanical coupling strength $g$ through 
$\sigma_{k,\text{out}}$, and is a function of the correlation matrix of the cavity optomechanical system. In our subsequent discussion, we analyze
Eq. \eqref{Wigner} in the context of a quantum estimation strategy based on the quantum Fisher information. Our task will be to seek optimal balanced homodyne photodetection measurement strategies.

\section{Quantum and classical Fisher Information}
\label{II}

In this section, we derive the quantum Fisher information (QFI) $H_g$ of the optomechanical coupling strength for a general Gaussian state, employing the phase-space description provided by the Wigner quasiprobability distribution 
\eqref{Wigner}. The QFI defines a lower bound for the mean-squared error (MSE) of an estimation setup, which is ensured by the quantum Cram\'{e}r--Rao theorem \cite{Helstrom2},
\begin{equation}\label{QCRB}
\text{MSE}(g)\geq\frac{|x'(g)|^2}{H_g}, 
\end{equation}
where $x'(g)$ is the derivative of the average estimator. When the estimator is unbiased, then $x'(g)=1$. The QFI is given as
\begin{equation}\label{eq:QFI}
H_g=\text{Tr}[\hat{\rho} \hat{L}_g^2],
\end{equation}
where $\hat{L}_g$ is the symmetric logarithmic derivative (SLD) defined by the equation 
\begin{equation}\label{SLD}
\partial_g\hat{\rho}=\frac{1}{2}\{\hat{\rho},\hat{L}_g\}.
\end{equation} 

We are going to use this general formalism to the Gaussian state obtained in Eq. \eqref{Wigner}. A Gaussian state is completely determined by its first and second moments;
however, here we have that the first moment is zero, following the argument in Sec. \ref{I}.  Since the density operator $\hat{\rho}$ of a Gaussian state can be expressed in an 
exponential form \cite{Adam}, we can write the operator $\hat{L}_g$ as a function of the covariance matrix $\sigma_{k,\text{out}}$. We neglect all subscripts in the subsequent discussion because, from now on, 
we focus on one mode of the electromagnetic field that is detected. 

In order to find the SLD, we use the Weyl transform on the operator, obtaining
\begin{equation}\label{WTSLD}
L(x,p)=\xi^T\Phi\xi-\nu,
\end{equation} 

\noindent where the explicit forms of $\Phi$ and $\nu$ are

\begin{eqnarray}\label{eq:Phi}
\Phi & = & -\frac{1}{2}\partial_g{(\sigma^{-1})}\\
\nu & = & -\frac{1}{2}\partial_g\ln(\det{\sigma}) = \text{Tr}[\Phi \sigma].
\end{eqnarray}
It is worth noticing that the quadratic nature of $L(x,p)$ is ensured by the Gaussian form of $W(x,p)$.

We use the SLD to calculate the QFI $H_g$ related to the parameter $g$ following \eref{QFI}. 
However, the calculation of the Weyl transform of $\hat{L}_g^2$ is not straightforward. In order to calculate it, we need to Weyl transform the function $L(x,p)$ 
back to the operator $\hat{L}_g$, yielding
\begin{equation}
\hat{L}=\Phi_{11}\hat{x}^2+\Phi_{12}(\hat{x}\hat{p}+\hat{p}\hat{x})+\Phi_{22}\hat{p}^2-\nu\mathds{1}.
\end{equation}
Now, one is able to calculate $\hat{L}_g^2$ and after performing the symmetric ordering and the Weyl transform on it, we find
$L^{(2)}(x,p)$ as
\begin{eqnarray}\label{L2Wigner}
L^{2}(x,p) & = & \Phi_{11}^2 x^4+4\Phi_{11}\Phi_{12}px^3+4\Phi_{12}^2p^2x^2\nonumber\\
&& +2\Phi_{11}\Phi_{22}p^2x^2+4\Phi_{12}\Phi_{22}p^3x+\Phi_{22}^2p^4\nonumber\\
&& -2\nu L(x,p)-\nu^2-\frac{1}{2}\det(\Phi).
\end{eqnarray}
The QFI obtained as the mean value of $\hat{L}^2$ on the state $\hat{\rho}$ can be calculated by the phase-space formalism,
\begin{eqnarray}\label{eq:QFI2}
H(g) & = & \int dxdp \, L^{(2)}(x,p)W(x,p)\nonumber\\
& = & 3\text{Tr}[(\Phi\sigma)^2]-2\nu\text{Tr}[\Phi\sigma]+2\det(\sigma)\det(\Phi)\nonumber\\
& & -\frac{1}{2}\det(\Phi)+\nu^2.
\end{eqnarray}

We will make use of \eref{QFI2} to determine the QFI of the output field. Combining together \eqref{eq:QFI2} and \eqref{eq:Phi}, we obtain the QFI for a two-dimensional Gaussian state with zero mean,
\begin{eqnarray}\label{eq:QFIdelta02}
H(g)&=&\frac{1}{2}\text{Tr}[(\partial_g(\sigma^{-1})\sigma)^2]-\frac{1}{8}\det[\partial_g(\sigma^{-1})].
\end{eqnarray}

The quantum Cram\'{e}r--Rao bound \eqref{QCRB} for an unbiased estimator is saturated only if we implement the best strategy (POVM) that minimizes the MSE of the parameter estimation. 
This strategy is usually very difficult to find and may be impossible to implement \cite{Bernad1}. However, we can find, for each practical measurement 
strategy, the maximum amount of Fisher information it can provide. Measurements on quantum systems provide a probability density function which 
depends on the parameter to be estimated. The amount of information on the unknown parameter carried by this probability density function can be measured by the so-called 
classical Fisher information (CFI)
\begin{equation}\label{eq:CFI}
F_g=\int dx P(x;g)\bigg{(}\partial_g\ln P(x;g)\bigg{)}^2,
\end{equation}
where $P(x;g)$ is the probability of obtaining the output of the measure $x$ when the true value of the parameter is $g$. In quantum mechanics, this probability is given by the relation $P(x;g)=Tr[\hat{\rho}_g\Pi_x]$. Here, we consider that the measurements are performed by balanced homodyne photodetection (BHD) \cite{Raymer}. 
This makes use of two photodetectors, each with quantum efficiency $\eta$. In BHD, the data recorded are proportional to the difference of the measured photon numbers $n_{1,2}$ of the two 
photodetectors, yielding
\begin{eqnarray}\label{HomodyneProb}
P_\theta(n;g) & = & \sum_{m=0}^{\infty}\langle :e^{-\eta(\hat{n}_1+\hat{n}_2)}\frac{(\eta\hat{n}_1)^{n+m}}{(n+m)!}\frac{(\eta\hat{n}_2)^m}{m!}: \rangle_{\hat{\rho}_g\otimes \hat{\rho}_{\text{LO}}}\nonumber
\end{eqnarray}
where $\hat{\rho}_{LO}$ is the state of the local oscillator, considered to be a coherent state $\hat{\rho}_{LO}=|\alpha_{LO}\rangle\langle\alpha_{LO}|$. The symmetric order denoted by $:\,:$ 
helps us to find the Weyl transform of the element $\Pi_k^{\eta}(x,p)$ of the BHD POVM, 
\begin{equation}\label{eq:POVMhomodynetransform}
\Pi_k^{\eta}(x,p)=\exp\Bigl[-\frac{2\eta(k-\frac{x\cos\theta+p\sin\theta}{\sqrt{2}})^2}{1-\eta}\Bigr]
\end{equation}
obtained in the limit of $|\alpha_{\text{LO}}|\gg 1$. The parameter $\theta$ is the angle of the coherent state $\ket{\alpha_{\text{LO}}}$, i.e., $\theta=\arg(\alpha_{\text{LO}})$, and defines the quadrature that 
is measured. The probability is
\begin{equation}\label{eq:HomodyneProb3}
P_\theta^\eta(k;g)=\sqrt{\frac{2\eta}{\pi(1-\eta)}}\int dx dp W(x,p)\Pi_k^{\eta}(x,p).
\end{equation} 
Using the condition \eqref{Wigner} for the Wigner function leads to
\begin{eqnarray}\label{eq:Pthetaetak}
P_{\theta}^{\eta}(k;g) & = & \frac{1}{\pi}\sqrt{\frac{2\eta}{1-\eta+2\eta R_\theta^T\sigma R_\theta}}e^{-\frac{2\eta k^2}{1-\eta+2\eta R_\theta^T\sigma R_\theta}}\nonumber,
\end{eqnarray}
where we have defined $R_\theta = (\cos\theta,\sin\theta)^T$. Now, we can calculate the CFI with the help of \eref{CFI}, yielding
\begin{eqnarray}\label{eq:CFIBHD}
F_g^\eta & = & 2\bigg{(}\frac{\eta R_\theta^T\partial_g\sigma R_\theta}{1-\eta+2\eta R_\theta^T\sigma R_\theta}\bigg{)}^2.
\end{eqnarray}
Equation \eref{CFIBHD} is a compact form for the classical Fisher information of the BHD. Notice that in the case of perfect detectors ($\eta\to1$) \eref{POVMhomodynetransform} reduces to a Dirac delta $\Pi_k^1(x,p)\to\delta(k-R^T\xi)$.
In this case, the CFI assumes the form
\begin{equation}
F_g^1=\bigg{(}\frac{R_\theta^T\partial_g\sigma R_\theta}{R_\theta^T\sigma R_\theta}\bigg{)}^2. \label{eq:CFIneed}
\end{equation}

\section{Results}
\label{III}

\begin{figure}
\includegraphics[width=.39\textwidth]{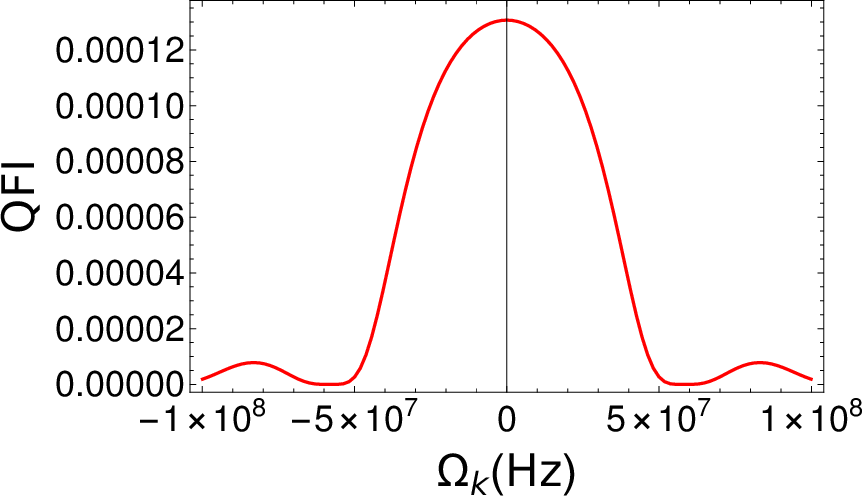}
\caption{Quantum Fisher information (QFI) as a function of the filter function's center frequency $\Omega_k$. 
In the rotating frame the peak of QFI is at $\Omega_k=0$, which corresponds to maximum accuracy reached in the estimation scenario. 
This also means that in the laboratory frame measurements should be performed around the frequency of the driving laser $\omega_{\text{L}}$. 
The parameters are $\kappa/2 \pi=18.5$ MHz, $\gamma/2 \pi=130$ Hz, $g/2 \pi=129$ Hz, $\omega_{\text{m}}/2 \pi=1.14$ MHz, $m=16$ ng, $T=11$ K, $\Delta_0=-2\kappa$ and the cutoff frequency 
$\Omega=5 \omega_{\text{m}}$. }
\label{fig:QFIDetectFreq}
\end{figure}

In Sec.~\ref{I}, we have calculated the covariance matrix \eqref{correlationmatrixoutput1} of the output field escaping the cavity and characterized by the filter function 
\eqref{eq:filter}. In Sec.~\ref{II}, we have calculated the general form of the quantum Fisher information 
(QFI) of Gaussian states with zero mean, like the output field, and the classical Fisher information (CFI) of balanced homodyne photodetection (BHD) measurements. In this section, we numerically 
investigate and compare QFI and CFI for an experimentally feasible situation. We consider the cavity to possess equal internal and external decay rates, $\kappa_{\text{in}}=\kappa_{\text{loss}}=\kappa$, and our detector to stay 
on for a temporal window of length $\tau=1/\kappa$. For our numerical analysis, we take the experimental values from \cite{Rossi}, which are $\kappa/2 \pi=18.5$ MHz, $\gamma/2 \pi=130$ Hz, 
$\omega_{\text{m}}/2 \pi=1.14$ MHz, $T=11$ K, $m=16$ ng, and the power of the laser, $P=1$ $\mu$W. Although the subject is to estimate the optomechanical coupling strength, we still need to set a central value 
around which we conduct our investigations. The coupling strength $g$ in Eq.~\eqref{OptomechanicalInteraction} has the dimensions of $[\text{Hz}\cdot\text{m}^{-1}]$, whereas in the experimental community, it
is common to give the dimensions of $g$ in Hertz \cite{Aspelmeyer2014}. We solve this by carrying out the transformation $g\to g\sqrt{\frac{2m\omega_{\text{m}}}{\hbar}}$.
Now, this transformed value according to Ref. \cite{Rossi} is $g/2 \pi=129$ Hz. In order to work in the high-temperature limit $k_B T \gg \hbar \Omega \gg  \hbar \omega_{\text{m}}$, 
we set the cutoff frequency $\Omega$ in Eq. \eqref{Ohmicspectraldensity} to $5 \omega_{\text{m}}$. In our subsequent numerical investigations, the detuning $\Delta_0$ is always chosen in such a way 
that bistability of the mechanical oscillator is 
avoided \cite{Karuza}. As a next step, we need to understand which central frequency of the filter function $\Omega_k$ gives us the best accuracy on the estimation of the coupling strength $g$. 
Therefore, we have calculated QFI as a function of $\Omega_k$, which has a peak at $\Omega_k=0$, as shown in Fig. \ref{fig:QFIDetectFreq}. Since we are in the rotating frame, this means that our detector 
filter function peaks at
the laser frequency $\omega_{\text{L}}$. 
%%%%%%%%%%%%%%%%%%%%%%%%%%%%%%%%%%%%%%%%%%%%%%%%%%%%%%%%%%%%%%%%%%%%%%%%%%%%%%%%%%%%%%%%%%%%%%%%%%%%%%%%
%%%%%%%%%%%%%%%%%%%%%%%%%%%%%%%%%%%%%%%%%%%%%%%%%%%%%%%%%%%%%%%%%%%%%%%%%%%%%%%%%%%%%%%%%%%%%%%%%%%%

\begin{figure}
\includegraphics[width=.39\textwidth]{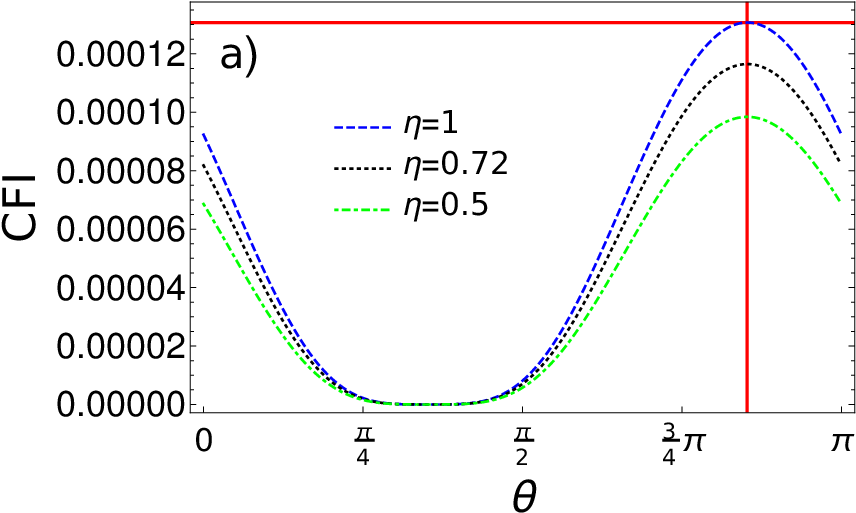}
\includegraphics[width=.39\textwidth]{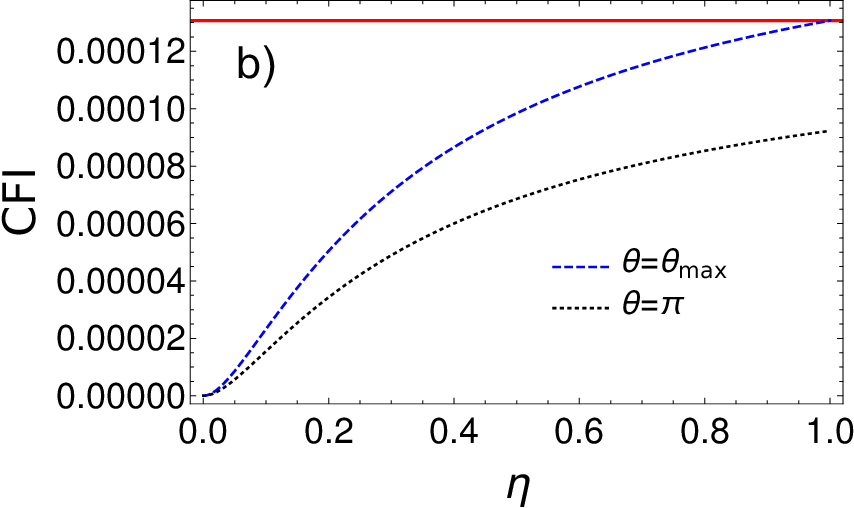}
\caption{a) Classical Fisher information (CFI) as a function of the local oscillator phase $\theta$ in the homodyne measurement for different values of $\eta$.  b) CFI as a function of $\eta$ for two  different choices of $\theta$. 
$\Omega_k=0$ and the rest of the parameters are the same of Fig. \ref{fig:QFIDetectFreq}.}
\label{fig:FigEtaTheta}
\end{figure}

Our goal is to find conditions for which the BHD results in the best achievable estimation strategy. This would correspond to the saturation of the quantum Cram\'{e}r--Rao bound. 
The outcome of the BHD depends on the quantum efficiency of the detectors $\eta$ and on the quadrature phase $\theta$ that we choose to measure. 
Figure \ref{fig:FigEtaTheta} shows the CFI as a function of these parameters. We notice that in the case of ideal detectors, i.e., $\eta=1$, the optimal choice for the phase 
$\theta=\theta_{\text{max}}$ leads the CFI to saturate the upper limit given by the QFI. This is a remarkable result that allows us to consider BHD as the optimal measurement that 
gives us the best estimate of $g$. In fact, Fig. \ref{fig:FigEtaTheta}(b) 
shows that the detector's efficiency $\eta$ is a very important parameter that affects the quality of the measurement, although it is no surprise that ideal photodetection results in
an optimal measurement scenario.

%%%%%%%%%%%%%%%%%%%%%%%%%%%%%%%%%%%%%%%%%%%%%%%%%%%%%%%%%%%%%%%%%%%%%%%%%%%%%%%%%%%%%%%%%%%%%%%%%%%%%%%%
%%%%%%%%%%%%%%%%%%%%%%%%%%%%%%%%%%%%%%%%%%%%%%%%%%%%%%%%%%%%%%%%%%%%%%%%%%%%%%%%%%%%%%%%%%%%%%%%%%%%

In general, an analytical solution for $\theta_{\text{max}}$ is very cumbersome because the global maximum of either Eq. \eqref{eq:CFIBHD} or Eq. \eqref{eq:CFIneed} depends on the entries of 
$\sigma$ in Eq.~\eqref{Wigner}, which are very complicated functions of the parameters of the system. However, the angle $\theta_{\text{max}}$ can be understood in the following way. 
Let us consider \eqref{eq:CFIneed}, which is 
the square of a generalized Rayleigh quotient for the self-adjoint matrix pairs $(\partial_g\sigma, \sigma)$ (see, for example, Ref. \cite{Horn}). Assuming that $\sigma$ is positive definite, i.e., 
does not describe a pure state, we would like to maximize
\begin{equation}
f(\theta)=\frac{R_\theta^T\partial_g\sigma R_\theta}{R_\theta^T\sigma R_\theta}. 
\end{equation}
The maximum value that $f(\theta)$ can reach is the maximum eigenvalue $\lambda_{\text{max}}$ of $\sigma^{-1/2}\partial_g\sigma\sigma^{-1/2}$ when $R_\theta$ is equal to the 
corresponding eigenvector $v_{\text{max}}$, which automatically defines $\theta_{\text{max}}$. In the case of \eqref{eq:CFIBHD}, we have a squared sum of generalized Rayleigh quotients \cite{Zhang}, and now,
$\theta_{\text{max}}$ can mostly be found by computational efforts. 
 
%%%%%%%%%%%%%%%%%%%%%%%%%%%%%%%%%%%%%%%%%%%%%%%%%%%%%%%%%%%%%%%%%%%%%%%%%%%%%%%%%%%%%%%%%%%%%%%%%%%%%%%%
\begin{figure}
\includegraphics[width=.39\textwidth]{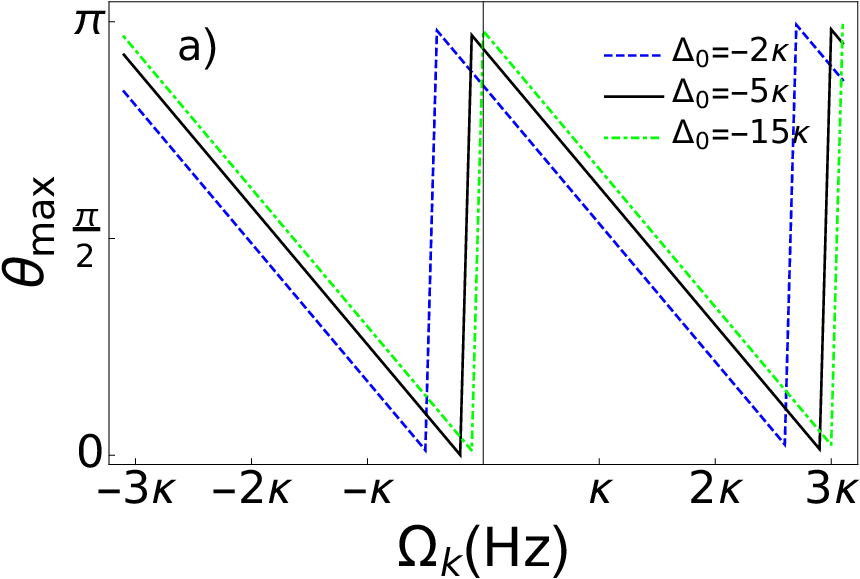}
\includegraphics[width=.41\textwidth]{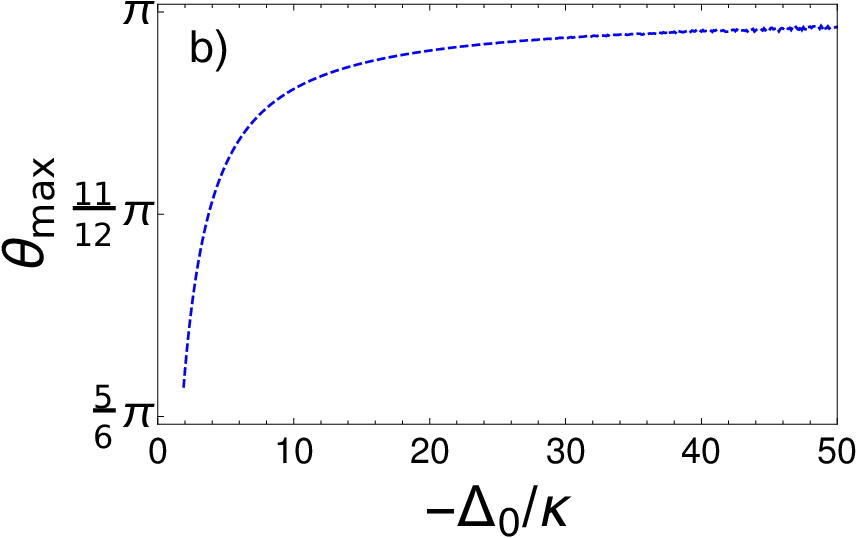}
\caption{ 
(a) The quadrature phase $\theta_{\text{max}}$ as a function of the filter function's center frequency $\Omega_k$ and for different values of the detuning $\Delta_0$. 
(b) $\theta_{\text{max}}$ as a function of the detuning $\Delta_0$ for $\Omega_k=0$. All other parameters are the same of Fig. \ref{fig:QFIDetectFreq}. }
\label{fig:FigTheta}
\end{figure}
%%%%%%%%%%%%%%%%%%%%%%%%%%%%%%%%%%%%%%%%%%%%%%%%%%%%%%%%%%%%%%%%%%%%%%%%%%%%%%%%%%%%%%%%%%%%%%%%%%%%

Our numerical investigations show that the value of $\theta_{\text{max}}$ is most sensitive to changes in the value of detuning $\Delta_0$ as well of the central frequency of the filter function $\Omega_k$. 
In order to gain some insight, we show in Fig. \ref{fig:FigTheta}(a) the dependence of $\theta_{\text{max}}$ with respect to $\Omega_k$. We can see that $\theta_{\text{max}}$ as a function of $\Omega_k$ follows an 
inverted ramp function. The ramp starts at $\kappa^2/\Delta_0$ and has a period of $3\kappa$. In the limit $\Delta_0\to-\infty$, $\theta_{\text{max}}$ tends to $\pi$. This is demonstrated in 
Fig.~\ref{fig:FigTheta}(b), which shows the value of $\theta_{max}$ as a function of the ratio $-\Delta_0/\kappa$. 

 %%%%%%%%%%%%%%%%%%%%%%%%%%%%%%%%%%%%%%%%%%%%%%%%%%%%%%%%%%%%%%%%%%%%%%%%%%%%%%%%%%%%%%%%%%%%%%%%%%%%%%%%
\begin{figure*}
\includegraphics[height=0.3\textwidth]{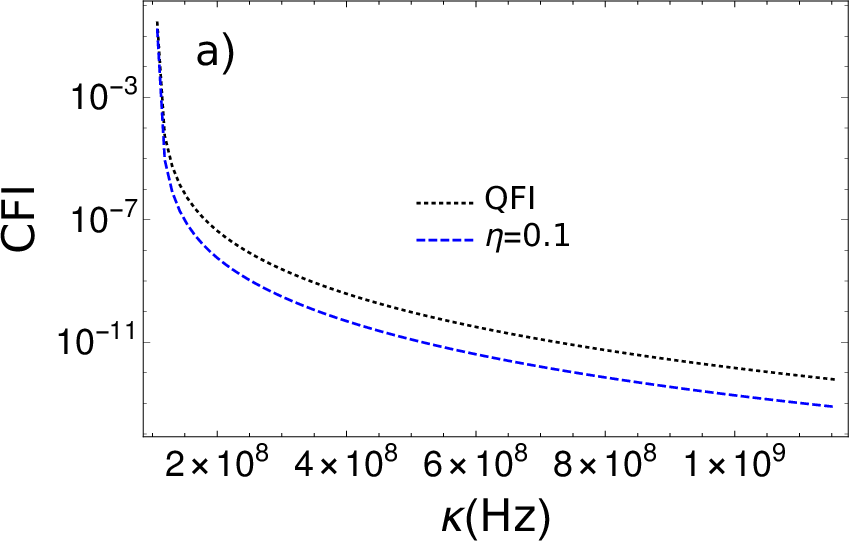}
\includegraphics[height=.3\textwidth]{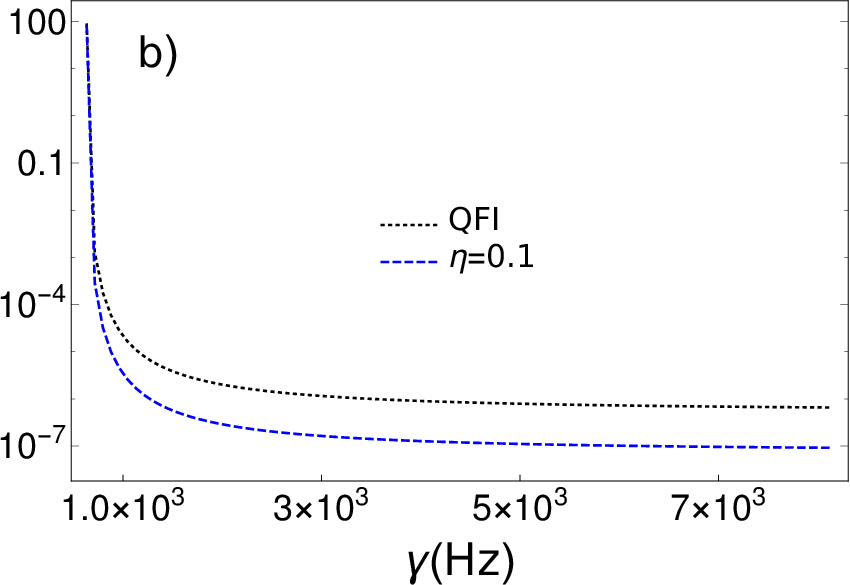}
\includegraphics[height=.3\textwidth]{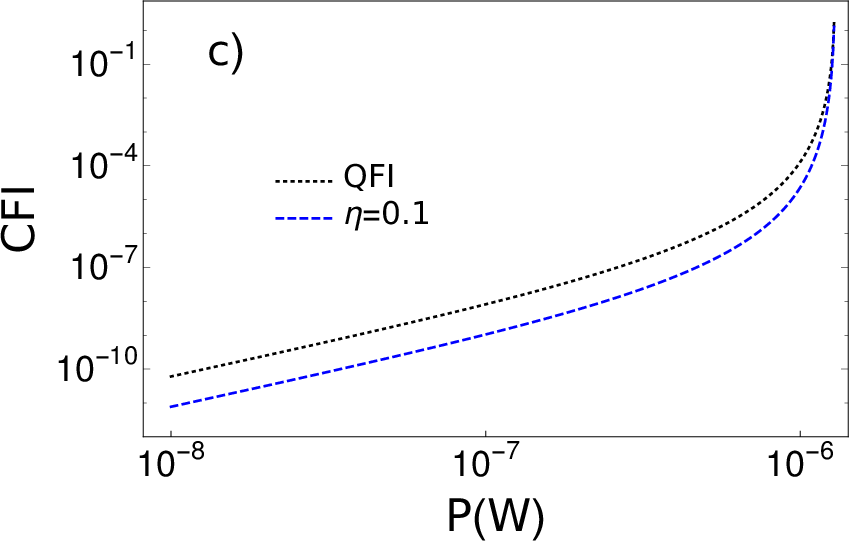}
\includegraphics[height=.3\textwidth]{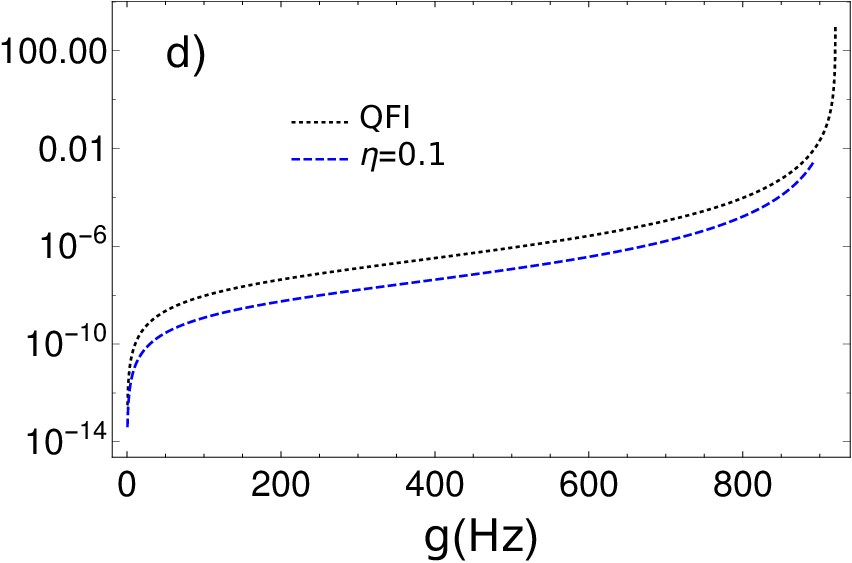}
\caption{Classical Fisher information (CFI) as the function of the parameters $\kappa$, $\gamma$, $P$ and $g$. We set $\Delta_0=-2\kappa$ and $\Omega_k=0$.
In the different figures we keep constant the parameter values of Fig. \ref{fig:QFIDetectFreq} and vary: a) the optical decay rate $\kappa$; b) the mechanical decay rate $\gamma$; c) the power of the 
driving laser $P$; and in d) 
the value of the coupling constant $g$.}
\label{fig:homodyneparameters}
\end{figure*}
%%%%%%%%%%%%%%%%%%%%%%%%%%%%%%%%%%%%%%%%%%%%%%%%%%%%%%%%%%%%%%%%%%%%%%%%%%%%%%%%%%%%%%%%%%%%%%%%%%%%

%%%%%%%%%%%%%%%%%%%%%%%%%%%%%%%%%%%%%%%%%%%%%%%%%%%%%%%%%%%%%%%%%%%%%%%%%%%%%%%%%%%%%%%%%%%%%%%%%%%%%%%%
\begin{figure}
\includegraphics[width=.39\textwidth]{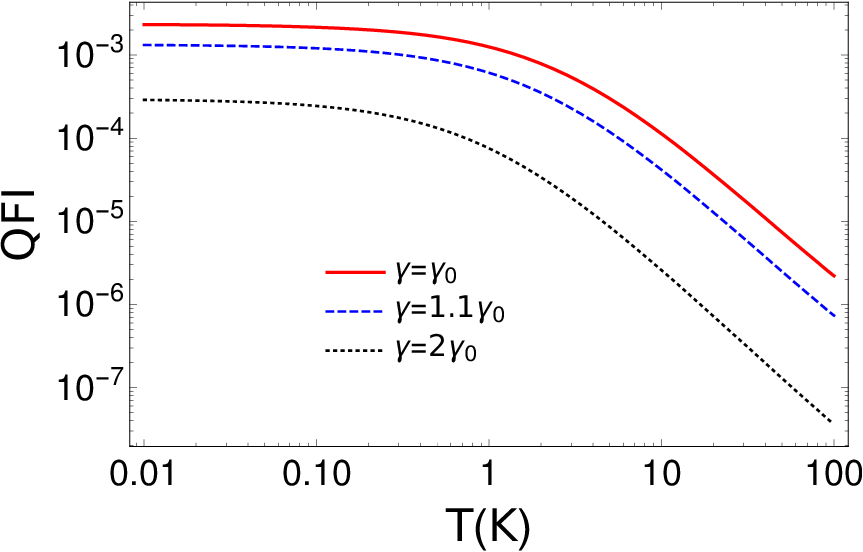}
\caption{Semilogarithmic plot of the quantum Fisher Information (QFI) as a function of the mechanical bath temperature $T$ for different values of the mechanical decay rate $\gamma$. 
Increasing  $\gamma$ leads to lower values of QFI. $\gamma_0/2 \pi=130$ Hz and the rest of the parameters are the same of Fig. \ref{fig:QFIDetectFreq}.}
\label{fig:QFITemp}
\end{figure}
%%%%%%%%%%%%%%%%%%%%%%%%%%%%%%%%%%%%%%%%%%%%%%%%%%%%%%%%%%%%%%%%%%%%%%%%%%%%%%%%%%%%%%%%%%%%%%%%%%%%

In the case of ideal detectors, i.e. $\eta=1$, and choosing $\theta=\theta_{\text{max}}$, the curve of CFI as a function of $\Omega_k$ perfectly overlaps with the curve of QFI in Fig. \ref{fig:QFIDetectFreq}. 
Thus, we can definitely set $\Omega_k=0$ for the rest of our numerical analysis because this choice guarantees the maximum accuracy in the estimation scenario. Now, we show the CFI, calculated numerically 
as we vary selected parameters that appear in the dynamical matrix $A$ of \eref{Amatrix}. The ranges of the plots are given by the stability conditions imposed on the system. 
Figures \ref{fig:homodyneparameters}(a) and \ref{fig:homodyneparameters}(b) show the CFI as a function of the optical 
and mechanical decay rate, respectively. These curves illustrate that increasing the decay rates lowers the accuracy of the estimation of the optomechanical coupling strength $g$. The opposite behavior is obtained 
increasing the power of the driving laser $P$; see Fig. \ref{fig:homodyneparameters}(c). In this case, a higher $P$ leads to a higher value for the stationary field amplitude $\lvert\alpha\rvert$, which leads to a more 
significant contribution to the dynamics from the optomechanical interaction as it appears in the interaction Hamiltonian  \eqref{OptomechanicalInteraction}.
Figure \ref{fig:homodyneparameters}(d) shows the CFI for different values of the optomechanical coupling strength. We remind the reader that the true value of $g$ is yet unknown. 
The CFI has its minimum for $g=0$, meaning that the accuracy is lower when the system
experiences weaker optomechanical interactions. Conversely, the CFI increases monotonically with the value of $g$ and it reaches its maximum when the system is on the threshold of the instability.

Figure \ref{fig:QFITemp} shows the QFI as a function of the temperature of the mechanical bath. For high temperature, the state tends to the maximally mixed state, regardless of the value of $g$, and the QFI decreases. 
For this reason, making measurements at low temperatures increases the estimation accuracy. QFI shows a sudden drop for a temperature around $1$ K. This value depends on the particular choice of the
spectral density \eqref{Ohmicspectraldensity} of the Brownian noise and on the other parameters of the system. As expected, increasing the value of $\gamma$ leads to lower values of QFI, and the accuracy of the 
estimation becomes worse.

\section{Conclusions}
\label{IV}

In this paper, we have investigated the estimation of the optomechanical coupling strength from the perspective of classical and quantum Fisher information. We have considered a cavity quantum electrodynamical
setup with a single mode of the electromagnetic field coupled to a single vibrational mode of a mechanical oscillator. Our model considers the quantum Brownian motion of the mirror and photon losses of the 
cavity field. We make use of the input--output formalism, motivated by the fact that experimental 
detection can be performed only on the output field of the cavity. The cavity is driven by a laser, which allows us to derive a set of linear quantum Langevin equations. Under these circumstances, we have been able 
to obtain the output field as a Gaussian state with zero mean as a stationary solution to the evolution of the whole system.

This Gaussian state as a function of the unknown optomechanical coupling strength determines the quantum Fisher information. Here, in contrast to the typical phase-space description (see Ref.~\cite{Monras}), 
our analysis makes an explicit use of the Weyl transform. To compare with experimentally relevant scenarios, we have considered a balanced homodyne photodetection strategy
as a realistic implementation for the estimation procedure. We have derived compact formulas for the quantum as well as the classical Fisher information.  

Finally, we have used the developed tools to investigate situations where the classical Fisher information is capable of saturating its upper limit, given by the quantum Fisher information. To make 
our findings more relevant, we have taken the experimental values of the system parameters from a recent study \cite{Rossi}. Our results show that the phase of the 
local oscillator in the balanced homodyne photodetection plays a crucial role because there are certain quadratures for which the classical Fisher information can reach the value of 
the quantum Fisher information. Under such conditions, the accuracy of the estimation, characterized by the mean-squared error, for certain quadrature measurements has the smallest lower bound.
Moreover, our investigation allows us to pinpoint the roles of the loss mechanism including less than ideal efficiency of the photodetectors. We have shown that the classical Fisher information is affected most 
strongly by nonideal detection efficiency. However, it is worth noticing that current state-of-the-art photodetectors have a close-to-ideal quantum efficiency \cite{Daiss}.

In conclusion, our analysis suggests that balanced homodyne photodetection plays a fundamentally important role in the estimation of the optomechanical coupling strength. However, it must be mentioned that our 
work is valid only for unbiased estimators. Furthermore, finding the smallest lower bound for the mean-squared error might reinforce our prior expectation of the value of the 
optomechanical coupling strength, as discussed in our earlier work \cite{Bernad2}. It is also important to mention that the use of $N$ independent and identical repetitions of the measurement scenario 
reduces the lower bound of the mean-squared error by a factor of $N^{-1}$ \cite{Helstrom}. Finally, our work bridges cavity optomechanics and inference techniques with the help of 
experimentally plausible models and parameter values and could serve as a guide for the optimal experimental characterization of optomechanical systems, which play important roles in gravitational 
waves detection \cite{Abbott}, but also in proposals for testing the conceptual bases of quantum mechanics \cite{Marshall, Adler, BDG, Carlesso}. 

\section{Acknowledgement}

This work is supported by the European Union’s Horizon 2020 research and innovation programme under 
Grant Agreement No. 732894 (FET Proactive HOT).

\appendix
\section{Dependence of the stationary field upon the laser power}
\label{sec:laser}

The condition imposed by Eq.~\eqref{conditionLaser} yields a third degree equation in the mean-field amplitude $\lvert\alpha\rvert$. The parameter $\varepsilon$ takes account of the number of photons inside 
the cavity and is related to the power $P$ of the laser as $\varepsilon=\sqrt{\frac{2 \kappa P}{\hbar\omega_{\text{L}}}}$. Equation \eqref{conditionLaser} can be written as a function of $P$ as
\begin{equation}\label{powerandamplitude}
P=\frac{|\alpha|^6 g^4 \hbar^3 \omega_{\text{L}}}{2\kappa m^2\omega_{\text{m}}^4}+\frac{|\alpha|^4\Delta_0 g^2 \hbar^2\omega_{\text{L}}}{\kappa m \omega_{\text{m}}^2}+\bigg{(}\frac{\kappa \omega_{\text{L}}\hbar}{8}+\frac{\Delta_0^2\hbar 
\omega_{\text{L}}}{2\kappa}\bigg{)}|\alpha|^2
\end{equation}
As is typical for non-linear systems, these equations exhibit bistability. This can be seen by putting together Eqs. \eqref{powerandamplitude} and \eqref{conditionshift}, expressing the shift of the mirror as a function of the laser power $P$, resulting in
\begin{equation}
q^{\text{(i)}}_0=\frac{\hbar g}{m\omega_{\text{m}}^2} A_{\text{i}},\quad i=1,2,3,\nonumber
\end{equation}
with $A_i=|\alpha|^2_i$ being the solution of \eqref{powerandamplitude}. The three curves intersect when $P=P_\pm$,
\begin{eqnarray}\label{powerlaser}
P_{\pm} & = & \frac{m \omega_{\text{L}} \omega_{\text{m}}^2 (-2\Delta_0(4\Delta_0+9\kappa^2)\pm\sqrt{(4 \Delta_0^2-3 \kappa ^2)^3})}{216 g^2 \kappa }\nonumber
\end{eqnarray}
In the regions $P<P_-$ and $P>P_+$, the system is stable, but for values in between, the system admits multiple solutions for $q_0$.
In this work, we considered a stable solution with a power $P<P_-$ and expressed the mean value as a function of the Hamiltonian parameters, $\alpha=\alpha(g,P,\Delta_0,\kappa)$.

\section{Covariance matrix of the output field}
\label{sec:sigmakout}

In this appendix, we derive the explicit form of the covariance matrix $\sigma_{k,\text{out}}$ of the filtered output field's quadrature $\hat{X}_{k,\text{out}},\hat{Y}_{k,\text{out}}$. 
We start from the filter function,
\begin{equation}
g_k(t)=\frac{\theta(t+\tau)-\theta(t)}{\sqrt{\tau}}e^{i\Omega_k t},\nonumber
\end{equation} 
where $\theta(t)$ is the Heaviside step function and $\tau$ is the temporal window when we detect the field. We consider the detector filter function to be peaked at frequency $\Omega_k$, with the 
condition that $\Omega_j-\Omega_l=\frac{2\pi}{\tau}n, n\in \mathbb{N}$.
The filter function is applied to the output field $\hat{a}_{\text{out}}$ in order to discretize the uncountably infinite number of modes forming the electromagnetic field outside the cavity into countably many modes. However, here we focus 
on a single mode detected by the measurement apparatus. This mode of the output field reads
\begin{equation}
\hat{a}_{k,\text{out}}(t)=\int_{-\infty}^t g_k(t-t')\hat{a}_{\text{out}}(t')dt'.\nonumber
\end{equation} 
Using the input--output relation $\hat{a}_{\text{out}}=\sqrt{\kappa}\hat{a}-\hat{a}_{\text{in}}$ we can write
\begin{equation}
\hat{a}_{k,\text{out}}(t)=\sqrt{\kappa}\int_{-\infty}^tdt'g_k(t-t')\hat{a}(t')-\int_{-\infty}^tdt'g_k(t-t')\hat{a}_{\text{in}}(t').\nonumber
\end{equation}
Hence, the quadratures $\hat{X}_{k,\text{out}},\hat{Y}_{k,\text{out}}$ read
\begin{widetext}
\begin{equation}
\hat{X}_{k,\text{out}}(t)=\int_{-\infty}^t ds\frac{\theta(t-s)-\theta(t-s-\tau)}{\sqrt{\tau}}\bigg{\{}\cos[\Omega_k(t-s)]\hat{X}_{\text{out}}(s)+\sin[\Omega_k(t-s)]\hat{Y}_{\text{out}}(s)\bigg{\}}\nonumber
\end{equation}
and, similarly,
\begin{equation}
\hat{Y}_{k,\text{out}}(t)=\int_{-\infty}^t ds\frac{\theta(t-s)-\theta(t-s-\tau)}{\sqrt{\tau}}\bigg{\{}\cos[\Omega_k(t-s)]\hat{Y}_{\text{out}}(s)-\sin[\Omega_k(t-s)]\hat{X}_{\text{out}}(s)\bigg{\}}.\nonumber
\end{equation}

We can write the covariance matrix as
\begin{eqnarray}
\sigma_{k,\text{out}}(t,s)&=&\kappa\int_{-\infty}^t\int_{-\infty}^s dt' ds' \frac{\theta(t'+\tau)-\theta(t')}{\sqrt{\tau}} G_k(t-t')\langle \hat{u}(t')\hat{u}(s')^T\rangle \frac{\theta(s'+\tau)-
\theta(s')}{\sqrt{\tau}}G_k(s-s')^T\\
&&+\int_{-\infty}^t\int_{-\infty}^s dt' ds' \frac{\theta(t'+\tau)-\theta(t')}{\sqrt{\tau}}G_k(t-t')\langle \hat{u}_{\text{in}}(t')\hat{u}_{\text{in}}(s')^T\rangle \frac{\theta(s'+\tau)-\theta(s')}
{\sqrt{\tau}}G_k(s-s')^T,\nonumber
\end{eqnarray}
\end{widetext}
where $\hat{u}(t)$ and $\hat{u}_{\text{in}}(t)$ have been defined in the main text and we have introduced the matrix
\begin{equation}
G(t)=\begin{pmatrix}
\cos{\Omega_kt} & \sin{\Omega_kt}\\
-\sin{\Omega_kt} & \cos{\Omega_kt}
\end{pmatrix}.\nonumber
\end{equation} 
Since we want to calculate the equal-time covariance matrix of the steady state, we can set $t=s$, take the limit $t\to\infty$, and substitute the Heaviside functions, yielding
\begin{eqnarray}\label{eq:sigmakout}
\sigma_{k,\text{out}}&=&\frac{\kappa}{\tau}\int_0^\tau\int_0^\tau dt' ds' G_k(t')\langle \hat{u}(t')\hat{u}(s')^T\rangle G_k(s')^T\nonumber\\
&+&\frac{1}{\tau}\int_0^\tau\int_0^\tau dt' ds' G_k(t')\langle \hat{u}_{\text{in}}(t')\hat{u}_{\text{in}}(s')^T\rangle G_k(s')^T\nonumber
\end{eqnarray}
We suppose the detector is switched on when the system has already reached the steady state. Furthermore, we consider the detection period $\tau$ to be small compared 
to the characteristic time of the steady state. This latter assumption has been tested with numerical simulations considering an integration time of the order of $\tau=\frac{1}{\kappa}$. 
In this scenario, we get
\begin{eqnarray}\label{eq:sigmakout2}
\sigma_{k,\text{out}}&=&\frac{\kappa}{\tau}\int_0^\tau\int_0^\tau dt' ds' G_k(t') \sigma G_k(s')^T\nonumber\\
&&+\frac{1}{\tau}\int_0^\tau dt'G_k(t')G_k(t')^T.
\end{eqnarray}
After performing the integrals in \eref{sigmakout2}, we obtain the expressions \eqref{correlationmatrixoutput1} to \eqref{correlationmatrixoutput3} of the main text.


\begin{thebibliography}{99}

\bibitem{Kaipio} J. Kaipio and E. Somersalo, {\it Statistical and Computational
Inverse Problems}, Applied Mathematical Sciences Vol. 1 (Springer-Verlag, New York, 2005).

\bibitem{Personick} S. D. Personick, IEEE Trans. Inf. Theory {\bf 17} (1971).

\bibitem{Holevo} A. Holevo, J. Multivar. Anal. {\bf 3}, 337 (1973).

\bibitem{Helstrom} C. W. Helstrom, {\it Quantum Detection and Estimation Theory} (Academic Press, New York, 1976).

\bibitem{Macieszczak} K. Macieszczak, M. Fraas, and R. Demkowicz-Dobrza\'{n}ski, New J. Phys. {\bf 16}, 113002 (2014).

\bibitem{Rzadkowski} W. Rzadkowski and R. Demkowicz-Dobrza\'{n}ski, Phys. Rev. A {\bf 96}, 032319 (2017).

\bibitem{Bernad1} J. Z. Bern\'{a}d, C. Sanavio, and A. Xuereb, Phys. Rev A {\bf 97}, 063821 (2018).

\bibitem{Rubio} J. Rubio and J. Dunningham, New J. Phys. {\bf 21}, 043037 (2019).

\bibitem{Bernad2} J. Z. Bern\'{a}d, C. Sanavio, and A. Xuereb, Phys. Rev. A {\bf 99}, 062106 (2019).

\bibitem{Aspelmeyer2014} M. Aspelmeyer, T. J. Kippenberg, and F. Marquardt, Rev. Mod. Phys. {\bf 86}, 1391 (2014).

\bibitem{Brawley2016} G. A. Brawley, M. R. Vanner, P. E. Larsen, S. Schmid, A. Boisen, and W. P. Bowen, Nat. Commun. {\bf 7}, 10988 (2016).

\bibitem{Bhattacharyya} A. Bhattacharyya, Sankh\={a}, Indian J. Stat. {\bf 8}, 1 (1946); {\bf 8}, 201 (1947).

\bibitem{Barankin} E. W. Barankin, Ann. Math. Statist. {\bf 20}, 477 (1949).

\bibitem{Rao} C. R. Rao, Bull. Calcutta Math. Soc. {\bf 37}, 81 (1945).

\bibitem{Cramer} H. Cram\'{e}r, {\it Mathematical Methods of Statistics} (Princeton University Press, Princeton, New Jersey, 1946).

\bibitem{Helstrom2} C. W. Helstrom, IEEE Trans. Inf. Theory {\bf 14}, 234 (1968).

\bibitem{Matteo} M. G. A. Paris, Int. J. Quantum Inf. {\bf 7}, 125 (2009).

\bibitem{Casella} G. Casella and R. L. Berger, {\it Statistical inference} (Duxbury, Pacific Grove, 2002).

\bibitem{Law} C. K. Law, Phys. Rev. A {\bf 49}, 433 (1994); {\bf 51}, 2537 (1995).

\bibitem{CL} A. O. Caldeira and A. J. Leggett, Physica {\bf 121A}, 587 (1983).

\bibitem{Breuer} H.-P. Breuer and F. Petruccione, {\it The theory of open quantum systems} (Oxford University Press, Oxford, 2002).

\bibitem{Walls} D. F. Walls and G. J. Milburn, {\it Quantum Optics} (Springer-Verlag, Berlin, 1994).

\bibitem{Raymer} A. I. Lvovsky and M. G. Raymer, Rev. Mod. Phys. {\bf 81}, 299 (2009).

\bibitem{Tagantsev} A. K. Tagantsev and S. A. Fedorov, Phys. Rev. Lett. {\bf 123}, 043602 (2019).

\bibitem{Breuer-Kappler} H.-P. Breuer, B. Kappler and F.~Petruccione, Ann. Phys. {\bf 291}, 36 (2001).

\bibitem{GiovVit} V. Giovannetti and D. Vitali, Phys. Rev. A {\bf 63}, 023812 (2001).

\bibitem{Garg} A. Garg, J. N. Onuchic, and V. Ambegaokar, J. Chem. Phys. {\bf 83}, 4491 (1985).

\bibitem{Stegun} M. Abramowitz and I. A. Stegun, {\it Handbook of Mathematical Functions} (Dover Publ., New-York, 1968).

\bibitem{Gant} F. R. Gantmacher, {\it Applications of the Theory of Matrices} (Wiley, New York, 1959).

\bibitem{Genes2008} C. Genes, A. Mari, P. Tombesi, and D. Vitali, Phys. Rev. A {\bf 78}, 032316 (2008).

\bibitem{Adam} G. Adam, J. Mod. Opt. {\bf 42}, 1311 (1995).

\bibitem{Rossi} M. Rossi, D. Mason, J. Chen, and A. Schliesser, Phys. Rev. Lett. {\bf 123}, 163601 (2019).

\bibitem{Karuza} M. Karuza, C. Biancofiore, M. Galassi,R. Natali, G. Di Giuseppe, P. Tombesi and D. Vitali, AIP Conference Proceedings {\bf 1}, 1363 (2011).

\bibitem{Horn} R. A Horn and C. R. Johnson, {\it Matrix Analysis} (Cambridge University Press, Cambridge UK, 1999).

\bibitem{Zhang} L.-H. Zhang, Comput. Optim. Appl. {\bf 54}, 111 (2013).

\bibitem{Monras} A. Monras, arXiv:1303.3682 (2013).

\bibitem{Daiss} S. Daiss, and S. Welte, and B. Hacker, and L. Li, and G. Rempe, Phys. Rev. Lett. {\bf 122}, 133603 (2019).

\bibitem{Abbott} B. P. Abbott et al. (LIGO Scientific Collaboration and Virgo Collaboration), Phys. Rev. Lett. {\bf 116}, 061102 (2016). 

\bibitem{Marshall} W. Marshall, C. Simon, R. Penrose, D. Bouwmeester, Phys. Rev. Lett. {\bf 91}, 130401 (2003).

\bibitem{Adler} S. L. Adler, A. Bassi, and E. Ippoliti, J. Phys. A {\bf 38}, 2715 (2005).

\bibitem{BDG} J. Z. Bern\'ad, L. Di\'osi, T. Geszti, Phys. Rev. Lett. {\bf 97}, 250404 (2006).
 
\bibitem{Carlesso} S. McMillen, M. Brunelli, M. Carlesso, A. Bassi, H. Ulbricht, M. G. A. Paris, and M. Paternostro, Phys. Rev. A {\bf 95}, 012132 (2017). 

\end{thebibliography}
\end{document}